\documentclass[twocolumn,10pt]{asme2e} 
\usepackage{graphicx}

\usepackage{amssymb}
\usepackage{caption}
\usepackage[figurename=FIGURE]{caption}

\usepackage{amsthm}
\usepackage{bm}
\usepackage{amsmath}

\confshortname{the ASME 2020 \\ Dynamic Systems and Control Conference}
\conffullname{DSCC 2020}

\confdate{4-7}
\confmonth{October}
\confyear{2020}
\confcity{Pittsburgh}
\confcountry{USA}
\papernum{DSCC2020-3315}

\title{Stability and Control of Chaplygin Beanies Coupled to a Platform through Nonholonomic Constraints}

\author{Blake Buchanan, Matthew Travers, Howie Choset 
    \affiliation{
	Robotics Institute\\
	School of Computer Science\\
	Carnegie Mellon University\\
	Pittsburgh, Pennsylvania 15213\\
    Email: blakeb,choset,mtravers@andrew.cmu.edu
    }
}

\author{Scott Kelly
    \affiliation{
	Department of Mechanical Engineering\\ and Engineering Science\\
	University of North Carolina\\
	Charlotte, North Carolina 28223\\
    Email: scott@kellyfish.net
    }
}

\begin{document}
\maketitle

\begin{abstract}
{\it Many multi-agent systems in nature are comprised of agents that interact with, and respond to, the dynamics of their environment. In this paper, we approach the study of such agent-environment interactions through the study of passively compliant vehicles coupled to their environment via simple nonholonomic constraints. We first consider a single passively compliant Chaplygin beanie atop a platform having translational compliance, introduce the reduced equations for the system using the notion of nonholonomic momentum, and provide proof for its stability under arbitrary deformations of the elastic element modeling its compliance. We then direct our focus to results concerning the frequency response and control of passive Chaplygin beanies under actuation of the platform, discuss rich dynamical features arising from periodic actuation, and develop rules by which control can be exerted to collect and disperse multiple passive vehicles. We then discuss how the latter of these results clarifies the extent to which stable behavior can be excited in the system through exogenous control.}
\end{abstract}



\section{INTRODUCTION}
\begin{figure}[t]
    \centering
    \includegraphics[width=0.4\textwidth]{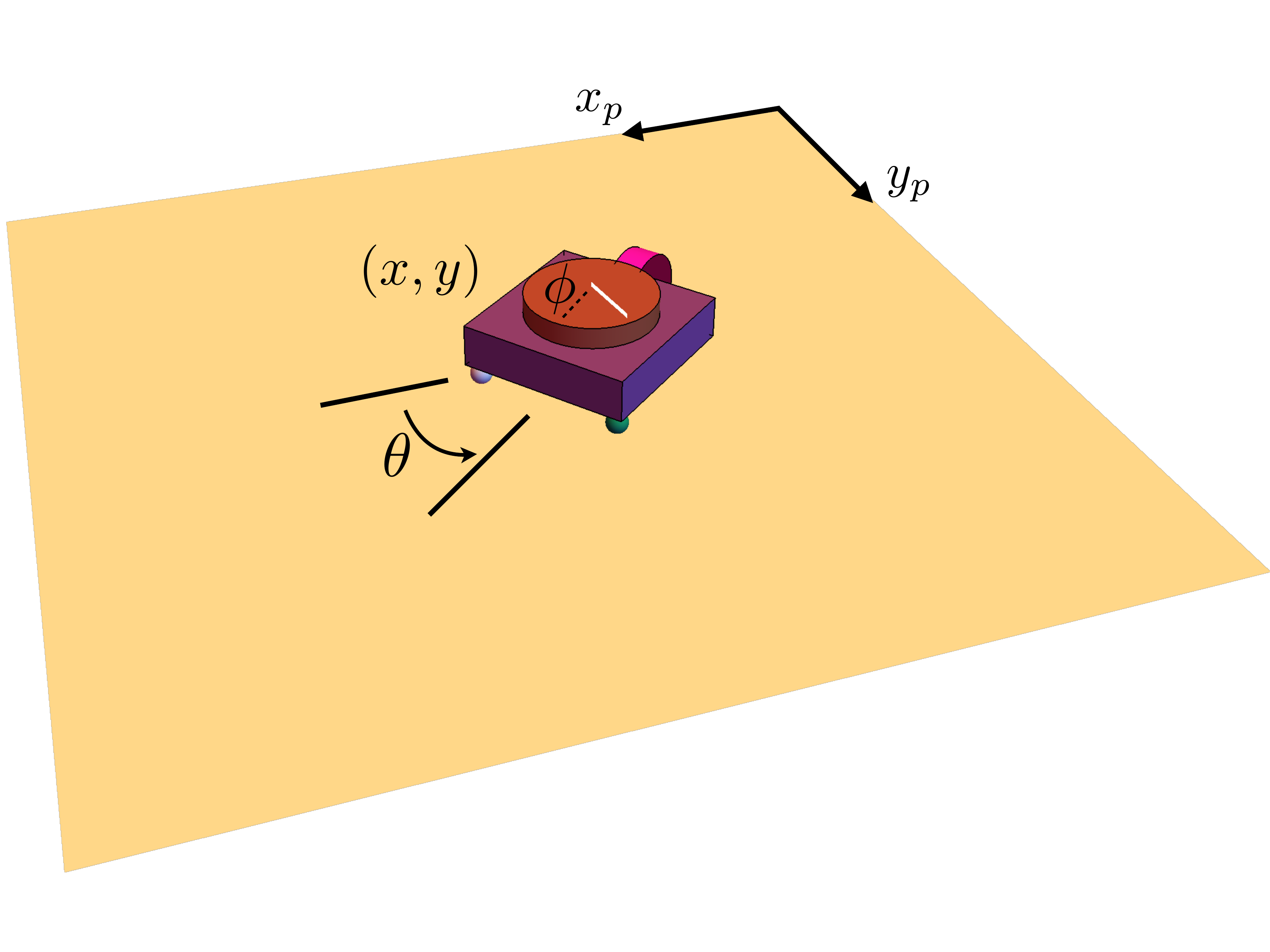}
    \caption{A Chaplygin beanie atop a translationally compliant platform. The vehicle's rotor angle relative to the heading is shown as $\phi$, its heading as $\theta$, its position relative to the platform as $(x,y)$, and the position of the platform in a laboratory frame, $(x_p,y_p)$.}
    \label{fig:fbd}
\end{figure}
In multi-agent robotic systems, it is not often that we consider the different and rich ways in which agents interact in their environment, especially when that environment possesses complicated dynamics of its own. In nature, however, there exist an abundance of systems which contain agents that move about in environments that respond dynamically to the locomotion of neighboring agents. Fish schools, bacterial swarms, and migratory cell groups are but a few, impacting their environment, say by shedding vortices in a fluid or pushing against surrounding compliant substrates, to effect motion. This particular class of systems exhibits the property that agent dynamics are \textit{coupled} through such substrates, motivating a deeper understanding of the mechanics underlying multi-agent coordination when considering the impact of substrate dynamics on agents. We ultimately seek to elucidate how an agent can exert control over its environment to both locomote in a desired way and to affect its neighbors in a way conducive to coordination. In this paper, we present a special case where we consider the scenario in which we assume direct control of the environment so as to induce locomotive behaviors of passive vehicles equipped with linear torsional springs, which we term \textit{exogenous control}.

We take the Chaplygin beanie\footnote{See \cite{kelly2012} for the appropriate etymology of the Chaplygin beanie.} as our motivating example --- effectively a Chaplygin sleigh with a rotor atop its body --- coupled to a translationally compliant platform through a nonholonomic constraint on its wheel, shown in Fig. \ref{fig:fbd}. We first consider the dynamics of the entire system to evolve only under its passive dynamics, i.e., there is no actuation in either the Chaplygin beanie or the platform atop which it sits. We develop the dynamics for this system using the method of nonholonomic reduction and prove that given a nonzero initial deformation in the spring coupling the vehicle's rotor to its body, the dynamics are asymptotically stable, with all of the system's angular momentum being converted into forward translational momentum. This stable behavior is likened to situations where a biological agent may prefer to relieve itself of actuation, taking advantage of its compliance and interactions with its environment to locomote.

We then assume control over the platform while allowing passively compliant Chaplygin beanies to deform under their own passive dynamics, modeled here as linear torsional springs. Specifically, we direct our focus to understanding the frequency response of passive Chaplygin beanies under exogenous periodic forcing. We characterize behaviors for the Chaplygin beanie as the actuation frequency of the platform is varied, provide bounds on this actuation, and develop rules by which control can be exerted to collect or disperse agents by exploiting knowledge of their physical characteristics.

In understanding motions of the platform to achieve particular behaviors in passive vehicles through exogenous control, we expect the problem of determining how agents should exert control over themselves in their environment to achieve collective behaviors in a multi-agent setting to become more approachable. Rich system dynamics are also encountered in our studies of this system that arise due to the nonholonomic constraints, suggesting the presence of multi-scale time dynamics.

\section{Prior Work}
Mechanical systems exhibiting nonholonomic constraints have recently been of utility in studying the effects of compliance in biological agents as well as the role of media coupling the dynamics of such agents. For example, systems like the \textit{Chaplygin beanie} \cite{kelly2012}, \textit{snakeboard} \cite{Bloch1996,shammas2012}, landfish \cite{dear2013}, and various \textit{nonholonomic snake robots} \cite{dear2016,dear2020}, have proven to be motivating examples in the control of biologically inspired robots. Specifically, the passive response many biological agents exhibit due to the natural compliance of joints or connective tissues has inspired the use of torsional springs to model compliance in mechanical systems with nonholonomic constraints \cite{Fedonyuk2019}. The utility in using reduced representations for proving the stability of such nonholonomic systems was demonstrated in \cite{kelly2012} and \cite{dear2013}. Additionally, recent works in multi-agent systems which are coupled to their environment have incorporated such compliant models \cite{kelly2018planar}. This coupling is seen predominantly in natural systems, e.g., schooling fish, swarming bacteria, or migrating cells \cite{Beal2006PassivePI, Darnton2010, Lo2000}, however, we provide evidence that an understanding of this coupling can be exploited to achieve meaningful behaviors for robotic systems as well.

The problem of exogenous control has been investigated in the context of transporting particles within a fluid at low Reynolds number using oscillating probes in \cite{ABRAJANGUERRERO2014}. Relatedly, \cite{Nitsan2016} used an oscillating probe to excite a substrate containing cardiac cells and showed that the induced deformations of the substrate due to exogenous forcing led to long-term oscillatory behavior in neighboring cardiac cells. A motion planning framework for robotic systems having external configurations, like those moving in dynamic environments, was established in \cite{dear2018}. 

\section{Equations of Motion}
We begin by developing a dynamic model for an entirely passive system, consisting of a passive vehicle atop a platform with finite inertia. The Chaplygin beanie will serve as our passive vehicle, equipped with a linear torsional spring coupling its rotor to its body. Initial displacements of the rotor relative to the body result in motion of both the passive vehicle and the platform atop which it sits. Motions of the platform in this case are due to the forces arising through the no-slip constraint at the wheel of the Chaplygin beanie.

\subsection{Modeling and Nonholonomic Reduction}
Constrained to the platform via a wheel located at its rear, the Chaplygin beanie locomotes using a rotor sitting atop  its body. The total mass of the vehicle is represented by $m$, its rotational inertia about the center of mass as $C$, rotational inertia of the rotor about the center of mass as $B$, and the mass of the platform as $M$. The distance between the center of mass and the contact point at the wheel is denoted by $a$, and the stiffness of the spring coupling the rotor to the body denoted by $k$. The position of the vehicle relative to the platform is given coordinates $(x, y)$, its orientation $\theta$, the rotor angle relative to the vehicle heading by $\phi$, and the position of the platform in a laboratory frame by $(x_p, y_p)$. The evolution equations arising from the reduction are the changes in linear and angular momentum permitted by the no-slip constraint at the wheel and will replace the equations describing the evolution of $\dot{x}$, $\dot{y}$, and $\dot{\theta}$. The presence of a platform gives rise to two additional evolution equations, one of which is the time evolution of forward translational momentum of both the Chaplygin beanie and the platform, the second of which is the time evolution of the momentum of both the Chaplygin beanie and the platform in the direction orthogonal to that allowed by the no-slip consraint at the wheel. We also refer to this momentum term as momentum \textit{lateral} to the forward motion of the vehicle. The Lagrangian for the system is given by
\begin{equation}
    \begin{aligned}
    L = & \frac{1}{2}m((\dot{x} + \dot{x}_p)^2 + (\dot{y} + \dot{y}_p)^2) + \frac{1}{2}C \dot{\theta}^2\\ + & \frac{1}{2} B(\dot{\theta} + \dot{\phi})^2
    + \frac{1}{2}M(\dot{x}_p^2 + \dot{y}_p^2) - \frac{1}{2}k \phi^2.
    \end{aligned}
\end{equation}
\noindent The nonholonomic constraint at the wheel is expressed as
\begin{equation}
    -\dot{x}\sin\theta + \dot{y}\cos\theta - a\dot{\theta} = 0.
\end{equation}
Constraints of this kind can also be thought of as one forms lying in the codistribution on the configuration manifold $Q$, written equivalently as
\begin{equation}
    \omega = -\sin\theta dx + \cos\theta dy - a d\theta.
\end{equation}
We require that the one form describing the no-slip constraint be annihilated by the system's generalized velocity, having coordinates $(x,y,\theta,\phi,x_p,y_p)$, at every point $q \in Q$. For the Chaplygin beanie on a platform with finite inertia, the manifold on which the dynamics evolve is described by the configuration manifold $Q = SE(2) \times \mathbb{S}^1 \times \mathbb{R}^2$.
\begin{figure}[t]
    \centering
    \includegraphics[width=0.45\textwidth,trim={1cm 4cm 0cm 4cm},clip]{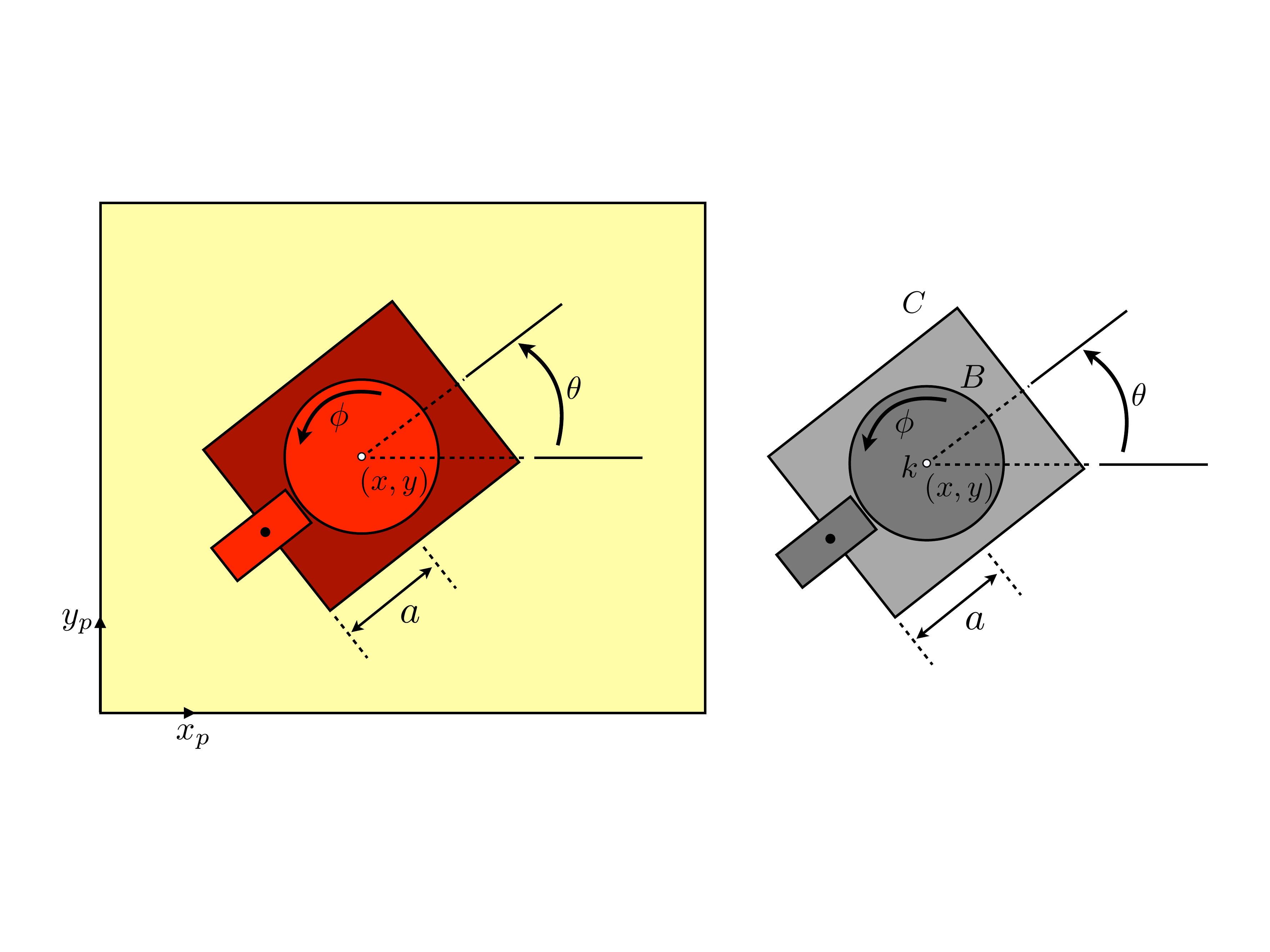}
    \caption{(Left) Two-dimensional diagram of a Chaplygin beanie on a platform with associated coordinates. (Right) Two-dimensional diagram of a Chaplygin beanie with parameter assignments.}
    \label{fig:fbd_single}
\end{figure}
Note that both $SE(2)$ and $\mathbb{R}^2$ together with matrix multiplication as the group operation are both Lie groups, and that their Cartesian product will also yield a Lie group with matrix multiplication as the group operation. The system's Lagrangian is invariant under the tangent lifted action, and the constraint one form is invariant under the cotangent lifted action, making $SE(2)\times \mathbb{R}^2$ a symmetry group. Thus, the Lie group $ G = SE(2) \times \mathbb{R}^2$ acts on the $G$ part of $Q$ via left translation, leaving the $\mathbb{S}^1$ part unchanged. 

We take an approach presented in \cite{Bloch1996}, involving the choosing of appropriate left-invariant vector fields spanning the intersection of the constraint distribution and the space tangent to the orbit of the group action, and leverage \cite{kelly2012} in computing the components of the nonholonomic momentum. We designate the distribution $\mathcal{D}_q$ as the space of all tangent vectors which annihilate the constraint one form, $\omega$, and let $T_q\text{Orb}(q)$ represent the space tangent to the orbit of the group action. We then choose appropriate vector fields on the configuration space, $Q$, to span
\begin{equation}
    S_q = \mathcal{D}_q \cap T_q\text{Orb}(q), \quad \forall q \in Q.
    \label{eqn:span}
\end{equation}
\noindent The following choice of vector fields is made.
\begin{equation}
    \begin{aligned}
    S_q & = \text{span}\{-a \sin \theta \frac{\partial}{\partial x} + a\cos\theta\frac{\partial}{\partial y} + \frac{\partial}{\partial \theta}, \\
    &\cos\theta \frac{\partial}{\partial x} + \sin\theta\frac{\partial}{\partial y}, \\
    & \cos\theta \frac{\partial}{\partial x_p} + \sin\theta \frac{\partial}{\partial y_p}, \\
    & -\sin\theta \frac{\partial}{\partial x_p} + \cos\theta\frac{\partial}{\partial y_p} \}.
    \end{aligned}
\end{equation}
The first two vector fields correspond to rotation about the contact point at the wheel and longitudinal translation of the vehicle, respectively \cite{kelly2012}. Flow along the third corresponds to forward translation of the entire system, including the platform, along the forward direction of the Chaplygin beanie. The fourth of these vector fields represents motions of the entire system lateral to the forward direction of the Chaplygin beanie. We invoke the Einstein summation convention in the following definition of the momentum map and designate $q^i$ as the $i$th coordinate on the configuration manifold, $Q$. The nonholonomic momenta are computed following 
\begin{equation}
    J^{nhc} = \frac{\partial L}{\partial \dot{q}^i} (\xi_Q)^i.
    \label{eqn:nonholomom}
\end{equation}
\twocolumn[
\begin{@twocolumnfalse}
\begin{equation}
    \begin{gathered}
        \dot{J}_{LT} = -\frac{m((B+C)J_Y + Ma(J_{RW} - B\alpha))(maJ_Y - (m+M)(J_{RW} - B\alpha))}{(M(ma^2 + B +C) + m(B+C))^2}, \\
        \dot{J}_{RW} = \frac{aJ_{LT}(maJ_Y - (m+M)(J_{RW}-B\alpha))}{M(ma^2 + B +C) + m(B+C)},  \\
        \dot{J}_X = -\frac{J_Y(maJ_Y +(m+M)(J_{RW}-B\alpha))}{M(ma^2 + B +C) + m(B+C)}, \\
        \dot{J}_Y = -\frac{J_X(-maJ_Y+(m+M)(J_{RW}-B\alpha))}{M(ma^2 + B +C) + m(B+C)} 
    \end{gathered} \label{eqn:evoeqns}
\end{equation}
\end{@twocolumnfalse}]
The resulting momenta are given directly by \eqref{eqn:nonholomom2}. It is clear by inspection that $J_{LT}$ and $J_{RW}$ correspond to forward translational momentum and angular momentum about the contact point of the wheel, respectively. The quantities $J_X$ and $J_Y$ correspond to forward translational momentum and momentum lateral to the direction allowed by the nonholonomic constraint for both the Chaplygin beanie and the platform.
\begin{equation}
    \begin{split}
      J_{LT} & = m(\dot{x} + \dot{x}_p)\cos\theta + m(\dot{y} + \dot{y}_p)\sin\theta, \\
        J_{RW} & = -ma(\dot{x} + \dot{x}_p)\sin\theta + ma(\dot{y} + \dot{y}_p)\cos\theta \\
        & + (B+C)\dot{\theta} + B\dot\phi,  \\
        J_{X}  = & m (\dot{x}+\dot{x}_p)\cos\theta + m(\dot{y}+\dot{y}_p)\sin\theta + \\
        & M\dot{x}_p\cos\theta + M\dot{y}_p\sin\theta,\\
      J_{Y}  = & -m (\dot{x}+\dot{x}_p)\sin\theta + m(\dot{y}+\dot{y}_p)\cos\theta - \\
        & M\dot{x}_p\sin\theta + M\dot{y}_p\cos\theta.
    \end{split} \label{eqn:nonholomom2}
\end{equation}

\begin{figure}[h]
    \centering
    \includegraphics[width=0.45\textwidth,trim={0 0cm 0 0cm},clip]{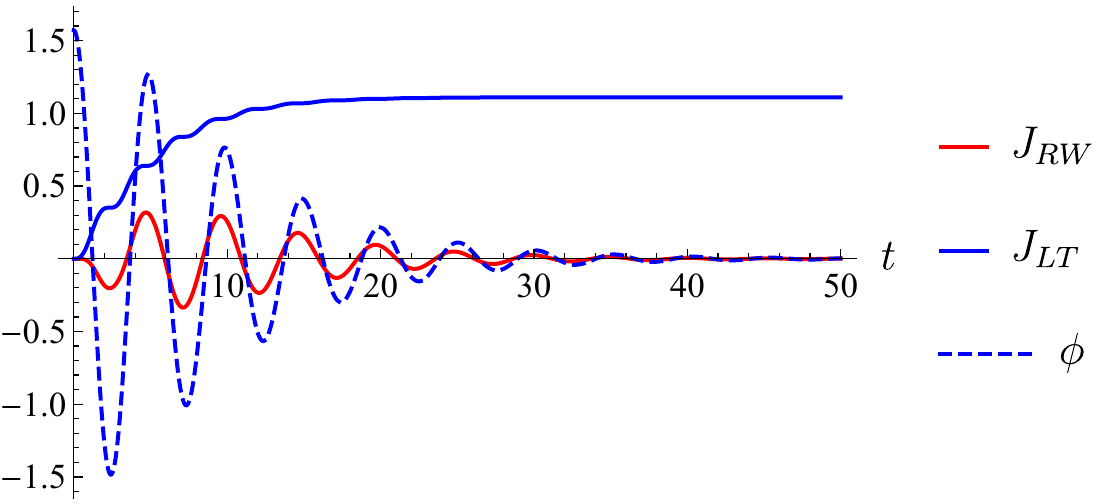}
    \caption{Rotational momentum about the rear wheel, longitudinal translational momentum, and the rotor angle}
    \label{fig:momentumplot}
\end{figure}

The evolution equations are computed following \eqref{eqn:nonholonomicevo} and are given by \eqref{eqn:evoeqns}, shown at the top of the page. 
\begin{equation}
    \dot{J}^{nhc} = \frac{\partial L}{\partial \dot{q}^i} \bigg[\frac{d\xi}{dt}\bigg]_Q^i
    \label{eqn:nonholonomicevo}
\end{equation}
The long-term behavior for this system under arbitrary initial $\phi$ is stable, with $J_{LT}$ tending to a constant value while $J_{RW}$ and $\phi$ tend to zero, as shown in Fig.
\ref{fig:momentumplot}.
\noindent In the following section, we present a formal argument for stable behaviors of this kind under the assumption that the system dynamics evolve on the zero level set of momentum, that is, any initial condition for which $J_X^2 + J_Y^2 = 0$.
\subsection{Stability}
Defining the following variables, the nonholonomic momenta, $\phi$, and $\dot{\phi}$ can be expressed as 
\begin{gather}
    r = \frac{J_{LT}}{d}, \quad w = \frac{J_{RW} - B\alpha}{d}, \quad p_x = \frac{J_X}{d},\label{eqn:neatermom} \\
    p_y = \frac{J_Y}{d}, \quad \alpha = \dot{\phi}.\nonumber
\end{gather}
The constants $d = m(B+C)+M(ma^2 + B + C)$, $\gamma_1 = -m^2a(B+C)/d$, $\gamma_2 = (m(m+M)(B+C)-m^2Ma^2)/d$, $\gamma_3 = mMa(m+M)/d$, $\lambda_1 = ma^2$, $\lambda_2 = -a(m+M)$, $\mu_1 = -ma$, $\mu_2 = m+M$, $\nu_0 = B/d$, $D = B(mMa^2 + C(m+M))$, $\nu_1 = -dk/D$, $\nu_2 = -Bma^2(m+M)/(dD)$, $\nu_3 = aB(m+M)^2/(dD)$, $\nu_4 = Bm^2a^2/(dD)$, and $\nu_5 = -mBa(m+M)/(dD)$ fully encapsulate the presence of system parameters in a more convenient form for analysis. The reduced dynamics are then easier to analyze for stability. Taking the time derivatives of \eqref{eqn:neatermom} and using \eqref{eqn:evoeqns} to make the necessary substitutions, the evolution equations become
\begin{equation}
    \begin{aligned}
        \dot{r}  &= \gamma_1 p_y^2 + \gamma_2p_yw + \gamma_3 w^2, \\
        \dot{w}  &=  \lambda_1 r p_y + \lambda_2 r w \\ & - \nu_0 (\nu_1 \phi + \nu_2 r p_y + \nu_3 r w + \nu_4 p_x p_y + \nu_5 p_x w),\\
        \dot{p}_x &= \mu_1p_y^2 + \mu_2p_yw, \label{eqn:neaterdyn}\\
        \dot{p}_y &= -\mu_1p_xp_y - \mu_2p_xw, \\
        \alpha &= \dot{\phi},\\
        \dot{\alpha} &= \nu_1 \phi + \nu_2 r p_y + \nu_3 r w + \nu_4 p_x p_y + \nu_5 p_x w.
    \end{aligned}
\end{equation}
The dynamics given by \eqref{eqn:neaterdyn} can be further simplified under assumptions of momentum conservation. The quantity $p_x^2 + p_y^2$ is conserved, with all its level sets invariant under \eqref{eqn:neaterdyn}. We wish to prove that all trajectories corresponding to $p_x^2 + p_y^2 = 0$ approach the $r$ axis asymptotically. Restricting the dynamics to this level set, \eqref{eqn:neaterdyn} can be written as
\begin{equation}
    \begin{gathered}
        \dot{r} = \gamma_3 w^2, \quad \dot{w} = \lambda_2 r w  - \nu_0 (\nu_1 \phi + \nu_3 r w),\\
        \dot{p}_x = 0, \quad \dot{p}_y = 0, \quad \alpha = \dot{\phi},\label{eqn:neaterdyn2}\\
        \dot{\alpha} = \nu_1 \phi + \nu_3 r w.
    \end{gathered}
\end{equation}
The time evolution of $r$, $w$, $\phi$, and $\alpha$ then fully describe the behavior of the system. By inspection it is clear that $\dot{r}$ is nonnegative, and is positive where $w$ is nonzero. For $w=0$, $\dot{w}$ is nonzero for $\phi \neq 0$. It follows that $r$ will increase for all time given that $w \neq 0$ and $\phi \neq 0$. Thus, $r$ will increase for all time unless $w$, $\phi$, and $\alpha$, are zero for all time, requiring $r$ to increase unless the flow of the vector field corresponding to \eqref{eqn:neaterdyn2} is always on the $r$ axis. All fixed positive values of $r$ correspond to a linear dynamical system described by $\dot{w}$, $\alpha$, and $\dot{\alpha}$. For every such $r$, denoted by $r_c$, the dynamics are then
\begin{equation}
    \begin{bmatrix} \dot{w} \\ \alpha \\ \dot{\alpha} \end{bmatrix} = \begin{bmatrix} \lambda_2 r_c w  - \nu_0 (\nu_1 \phi + \nu_3 r_c w) \\ \dot{\phi} \\ \nu_1 \phi + \nu_3 r_c w \end{bmatrix}.
    \label{eqn:awayfromr}
\end{equation}
The Jacobian of \eqref{eqn:awayfromr} is
\begin{equation}
    A = \begin{bmatrix} \lambda_2 \nu_0 \nu_3 r_c & -\nu_0 \nu_1 & 0 \\
    0 & 0 & 1 \\
    \nu_3 r_c & \nu_1 & 0 \end{bmatrix}.
    \label{eqn:jacobianawayfromr}
\end{equation}
The eigenvalues of \eqref{eqn:jacobianawayfromr} at $(w,\phi,\alpha) = (0,0,0)$ correspond to the roots of a third order polynomial in $p$ with parameter-dependent coefficients, written as
\begin{equation}
    p^3 + (\nu_0\nu_3 r_c - \lambda_2 r_c)p^2 - \nu_1p + \lambda_2\nu_1r_c = 0.
\end{equation}

With $r_c > 0$, the polynomial above clearly has roots with all negative real part, showing that $w$, $\phi$, and $\alpha$ exponentially decrease to zero as $r$ increases. This result suggests a stable fixed point of \eqref{eqn:neaterdyn2} at $(r,w,\phi,\alpha) = (r_\infty,0,0,0)$ given knowledge of the asymptotic values of $w$, $\phi$, and $\alpha$. Since the dynamics are energy-conserving, the initial and final energies of the system must be equal, with the asymptotic value of $r$ for the case where the system is initially at rest calculated as
\begin{equation}
    r_\infty = \lim_{t \to \infty} r =  \frac{\phi(0)}{d} \sqrt{\frac{kmM}{m + M}}. \nonumber
    \label{eqn:rinf}
\end{equation}
It follows that the asymptotic value of $r$ can be determined for any initial conditions corresponding to the system being initially at rest. Of considerable note is the linear relationship between $r_\infty$ and the initial rotor angle relative to the body of the Chaplygin beanie, i.e., the initial spring deformation.
\begin{figure}[t]
    \centering
    \includegraphics[width=0.45\textwidth,trim={0 5cm 0 5cm},clip]{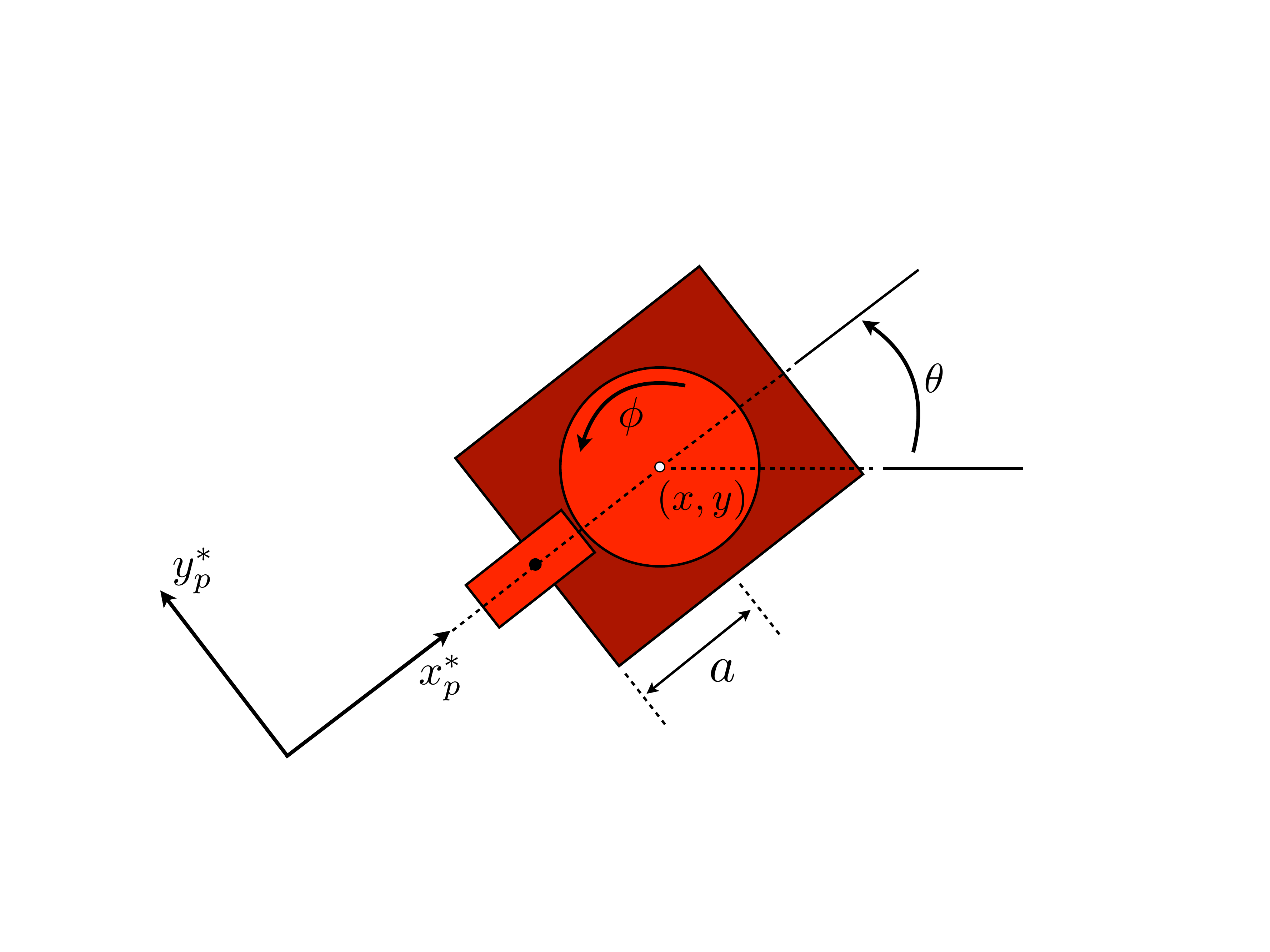}
    \caption{Platform actuation rotated so as to exert control in the direction orthogonal to direction of motion allowed by the no-slip constraint at the wheel}
    \label{fig:beaniePlatformActuation}
\end{figure}
\section{Exogenous Control}
\label{section:externalcontrol}
Coordination of biological agents in fluids or compliant substrates is often accomplished by the agent taking actions based on locally-sensed dynamics of their environment. Though agents take such actions, they also relieve themselves of their actuation under some circumstances, remaining passively compliant for some time in their environment before taking up actuation again. Such passivity has proven useful for achieving meaningful locomotive behaviors \cite{Beal2006PassivePI}. Prior works have considered this problem in the context of vibrational entrainment of passively compliant Chaplygin beanies atop an actuated platform \cite{kelly2018planar}. In this section, we study the Chaplygin beanie atop an actuated platform and ask whether the system can be excited so as to induce predictable locomotion. We investigate this problem from a perspective of exogenous control, i.e., we excite the platform in an oscillatory manner, uncaring of \textit{where} the control originates, and characterize motion primitives for a passively compliant Chaplygin beanie as it responds passively to its environment. 

\subsection{Frequency Response Analysis}
\label{subsection:frequency}
We seek to characterize locomotive behaviors when the frequency at which the environment is stimulated varies over a range of values containing the natural frequency of the rotor and the modal frequency of the body-rotor couple when not constrained to the platform, both of which are dependent on the stiffness of the spring. Consider a single passively compliant Chaplygin beanie atop an actuated platform and let its parameters, $m$, $B$, $C$, $a$, and $k$, be equal to unity. Note that the platform can only be actuated in the directions $(x_p,y_p)$ as shown in Fig. \ref{fig:fbd_single}. There's no reason to assume a relationship exists between forward translational speed, heading, or even stability, when actuating purely in the $(x_p,y_p)$ directions. In fact, such a relationship is obfuscated by dependence on the initial heading of the vehicle. However, such a relationship could exist when considering actuation in a rotated frame of reference, orthogonal to the allowable direction of motion required by the nonholonomic constraint at the vehicle's wheel. 

\begin{figure}[t]
    \centering
    \includegraphics[width=0.45\textwidth,trim={0 0cm 0 0cm},clip]{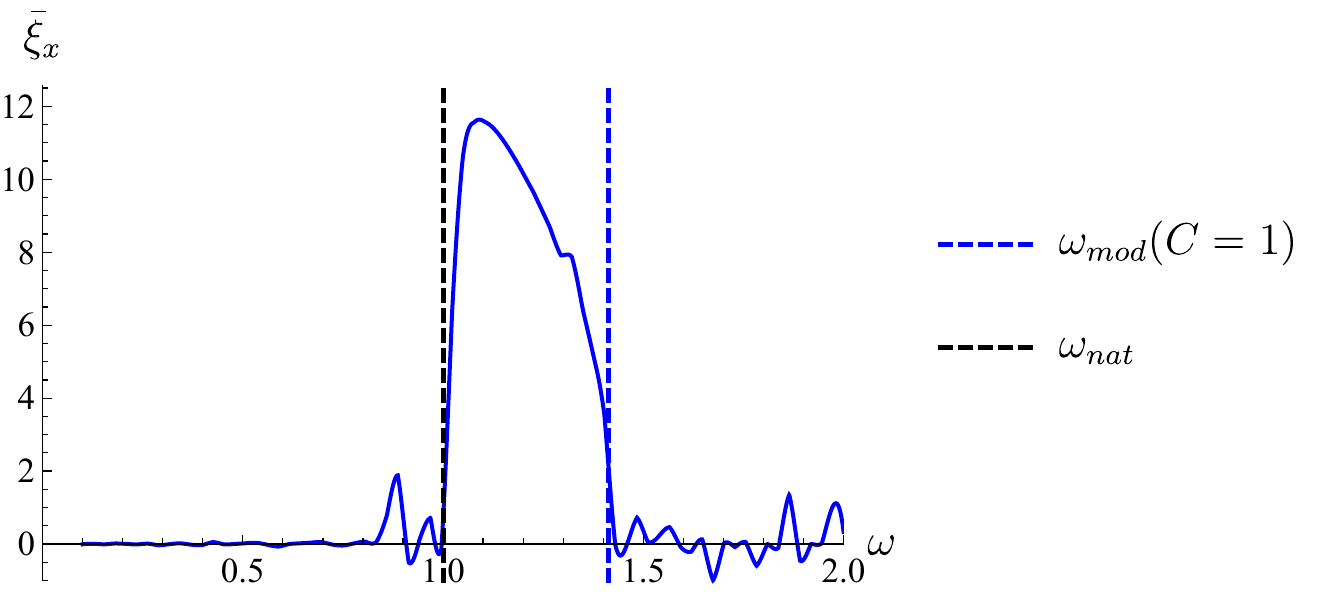}
    \caption{Frequency response of a passive Chaplygin beanie under external actuation in the body frame. The parameters were set to unity to obtain this response.}
    \label{fig:frequencyresponse}
\end{figure}

Fig. \ref{fig:beaniePlatformActuation} shows the rotated reference frame of the platform. Actuation along $y_p^*$ is not only independent of the heading of the vehicle, but also the direction for which its passive dynamics are most responsive. Actuation along $x_p^*$, for example, causes no deformations of the rotor relative to the body and therefore no passive response. Consider the natural frequency of the rotor and the modal frequency of the vehicle when not constrained to the platform, given by
\begin{equation}
    \begin{gathered}
        \omega_{nat} = \sqrt{\frac{k}{B}}, \qquad \omega_{mod}  = \sqrt{\frac{k(B+C)}{BC}}. \label{eqn:frequencies}\\
    \end{gathered} 
\end{equation}
With a spring coupling the cart to the rotor, we sweep through a range of frequencies for a particular set of parameters to analyze the response of the system to exogenous forcing in the $y_p^*$ direction and use asymptotic mean forward translational velocity as a performance metric. However, under certain periodic actuation the system will approach persistently undulatory behavior, requiring that we consider the mean velocity for an integer number of periods of oscillation. Forward translational velocity in the body frame of the vehicle is given by
\begin{equation*}
    \begin{aligned}
       \xi_x = \dot{x}\cos\theta + \dot{y}\sin\theta.
    \end{aligned}
\end{equation*}
\noindent The above is equivalent to the velocity in the direction allowable by the nonholonomic constraint on the wheel. Consider a situation that allows the orientation of the agent to be tracked in the environment. We actuate the platform according to
\begin{equation}
    \begin{aligned}
       \begin{bmatrix} x_p^* \\ y_p^* \end{bmatrix} = \begin{bmatrix}\cos\theta & -\sin\theta \\ \sin\theta & \cos\theta \end{bmatrix}^{-1} \begin{bmatrix} x_p \\ y_p \end{bmatrix}
       \label{eqn:controlrotated}
    \end{aligned}
\end{equation}
\noindent and let $x_p^* = 0$ and $y_p^* = A\sin(\omega t)$. Note that this is effectively a feedback-like controller in that it requires tracking the heading of the vehicle, which is then used to compute the control $y_p^*$. Setting system parameters and the amplitude $A$ to unity, we discretize the range of frequencies between 0.1 and 2.0 into $N=70$ equally-spaced intervals. Using the final three periods of oscillation to compute the mean forward translational velocity, denoted by $\Bar{\xi}_x$, we obtain the frequency response plot shown in Fig. \ref{fig:frequencyresponse}.

\begin{figure}[t]
    \centering
    \includegraphics[width=0.45\textwidth,trim={0 0cm 0 0cm},clip]{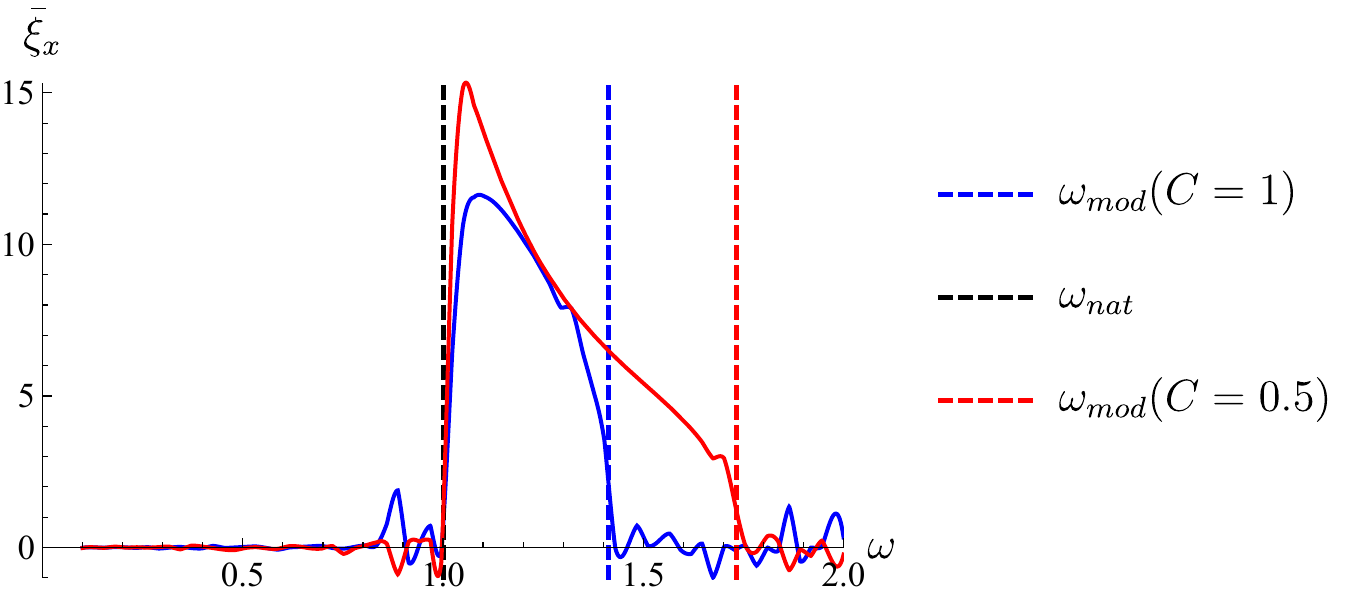}
    \caption{Frequency responses of a Chaplygin beanie for two different parameter combinations}
    \label{fig:frequencyresponse2}
\end{figure}

In carrying out this experiment, it is clear that the natural frequency of the rotor and modal frequency of the system in free space bound a region of high performance when considering mean forward translational velocity as a metric. Additional questions concerning generalizability and associated behaviors arise from this particular result. To address the first of these questions, we vary the parameters for the vehicle, compute the corresponding frequencies given in \eqref{eqn:frequencies} and generate similar results to Fig. \ref{fig:frequencyresponse}. The results of these experiments are shown in Fig. \ref{fig:frequencyresponse2}. The natural frequency of the rotor and modal frequency of the vehicle in free space again yield lower and upper bounds on regions of high performance.

Of further interest are behaviors emerging from actuating within, and outside of, the frequency bounds set by \eqref{eqn:frequencies}. In particular, we wish to characterize the frequencies that result in stable dynamics and from that characterization deduce motion primitives for controlling multiple passive vehicles.
\begin{figure}[t]
    \centering
    \includegraphics[width=0.45\textwidth,trim={0 0cm 0 0cm},clip]{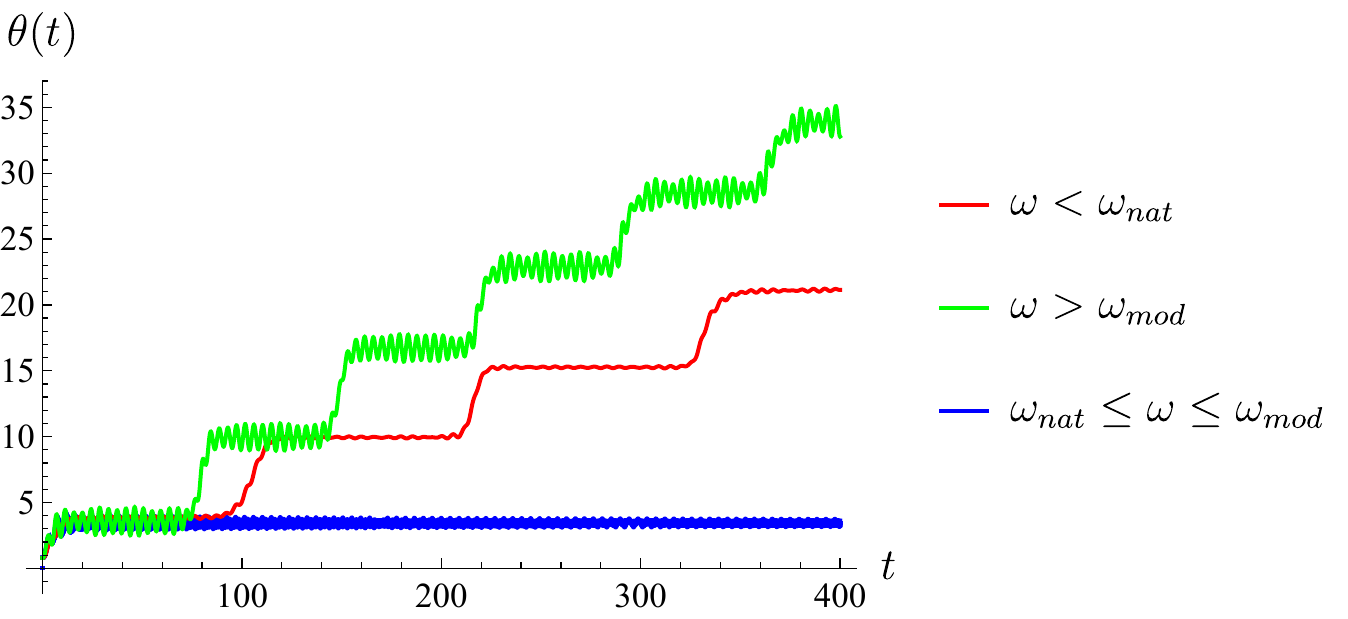}
    \caption{Analysis of the asymptotic heading of a Chaplygin beanie over the actuation bounds described in Fig. \ref{fig:frequencyresponse}}
    \label{fig:headinganalysis}
\end{figure}
An analysis of the time evolution of $\theta$ when actuating the platform at frequencies inside and oustide of the bands given above is shown in Fig. \ref{fig:headinganalysis}, clarifying the existence of distinct dynamics in the vehicle's heading. The blue curve visible in Fig. \ref{fig:headinganalysis} is actually a family of curves all resulting from frequencies lying within the bounds of the natural and modal frequency of the vehicle, all resembling stable oscillatory behavior. The red and green curves each correspond to $\theta$ dynamics of a single frequency chosen that satisfies the inequalities shown in the legend.

\subsection{Manipulation}
\begin{figure}[h]
    \centering
    \includegraphics[width=0.4\textwidth,trim={0 0cm 0 0cm},clip]{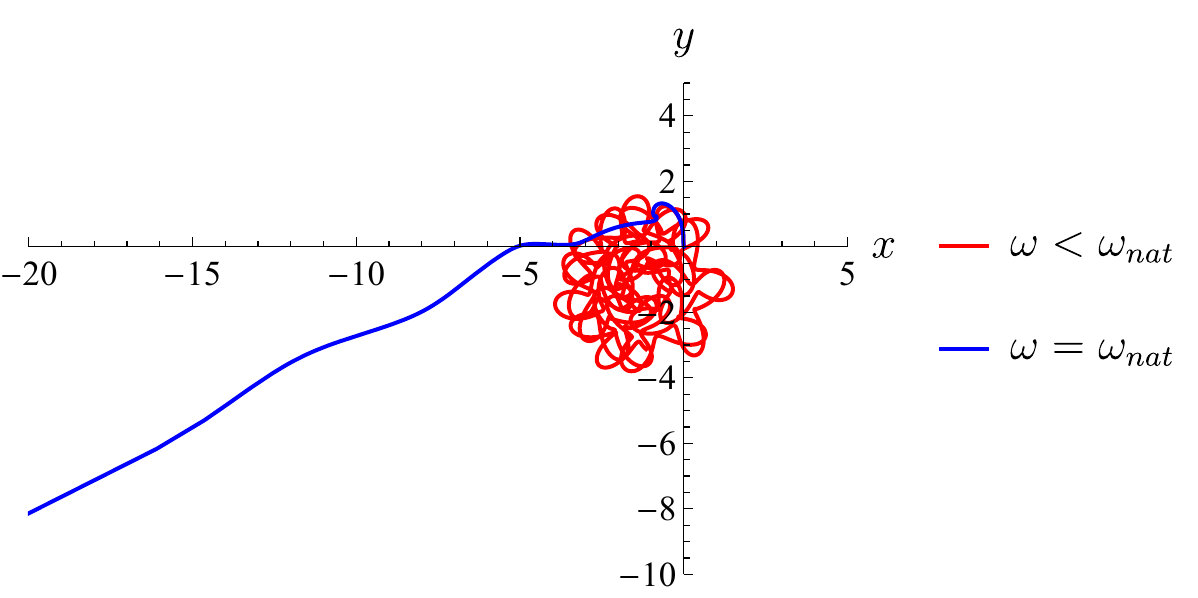}
    \caption{Trajectories of two individual simulations for actuation of the platform inside of the bounds (blue) and outside of bounds (red) defined by $\omega_{nat}$ and $\omega_{mod}$}
    \label{fig:trajectories}
\end{figure}

The frequency characterization in \ref{subsection:frequency} provides clear rules by which we can exert control over the platform to manipulate Chaplygin beanies. Actuating the platform at frequencies within the bounds set by the natural frequency of the rotor and the modal frequency of the vehicle in free space allow for control primitives which induce vehicles to achieve stable undulatory locomotion along a particular heading. We term such behavior in the context of manipulating multiple agents as \textit{dispersion}. Actuation outside of these boundaries yield trajectories corresponding to much more complex dynamics, not as easily classified as those of stable undulatory behavior. We discuss some of these behaviors in Section \ref{section:futurework}. Two such trajectories are shown in Fig. \ref{fig:trajectories}.

The beanie under external actuation with $\omega = \omega_{nat}$ will disperse from its initial position and undulate stably at a particular heading for all time. The degree to which it stably oscillates in $\theta$ increases with increasing $\omega$, as long as actuation stays within the bounds of the natural frequency of the rotor and the modal frequency of the vehicle. Though no formal guarantee is given in this work, the authors assert that trajectories corresponding to those of platform actuation at frequencies of $\omega < \omega_{nat}$ or $\omega > \omega_{mod}$ will stay within some neighborhood of its initial position, much like that of the red trajectory in Fig. \ref{fig:trajectories}. Such trajectories are also prone to exhibit dynamics that reveal the presence of multi-scale time dynamics, discussed below.

\begin{figure}[h]
    \centering
    \includegraphics[width=0.4\textwidth,trim={0 0cm 0 0cm},clip]{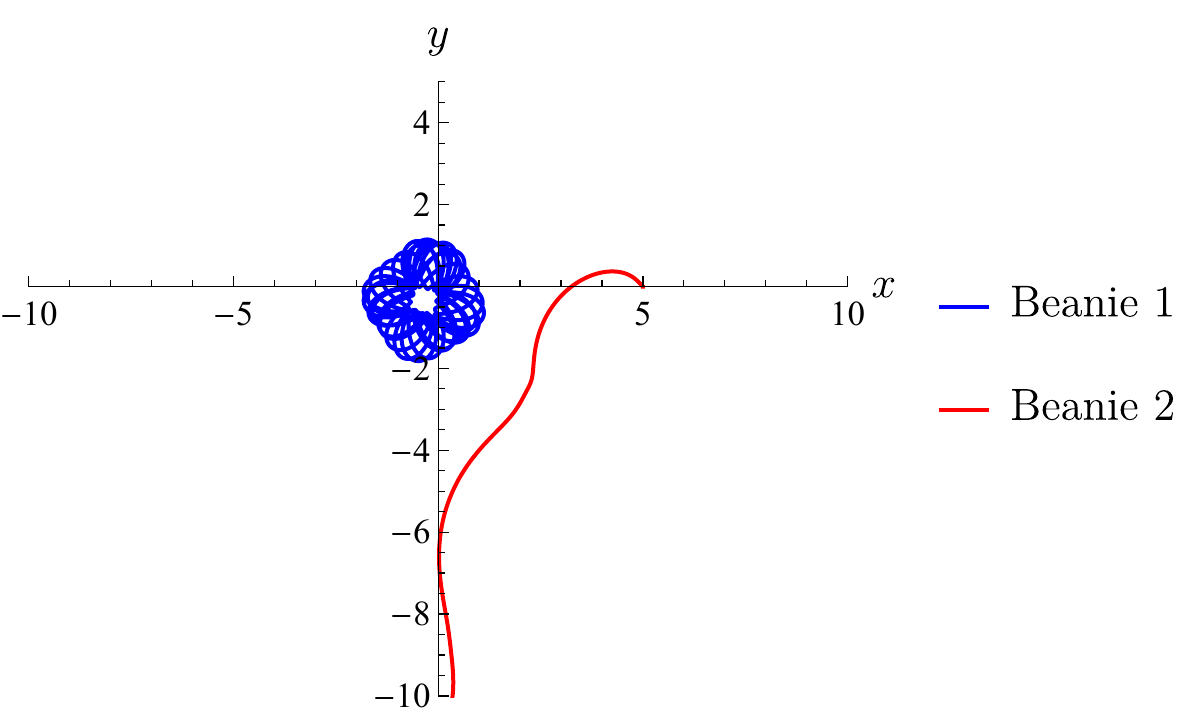}
    \caption{Resulting trajectories for two Chaplygin beanie agents atop an actuated platform. The blue trajectory corresponds to actuation of beanie 1 at a frequency lower than the natural frequency of its rotor. The red trajectory corresponds to the dynamics induced by actuating the platform so as to induce the behavior seen in beanie 1.}
    \label{fig:twoBeaniesOnAPlatform}
\end{figure}

This result clarifies the ability to control passive vehicles using the kind of actuation given by \eqref{eqn:controlrotated}. Consider a case with two identical passive Chaplygin beanies at rest atop an actuated platform with different initial headings. Naively assuming control over the platform to manipulate one vehicle in this sense does not guarantee a certain behavior for the other. In the presence of other passive vehicles, however, to achieve a desired behavior, the platform can be actuated corresponding to the desired control for that particular vehicle and its resulting behavior remains independent of the others. Fig. \ref{fig:twoBeaniesOnAPlatform} shows such a case if the desired behavior for beanie 1 is to stay within some neighborhood of its initial position.

Though beanie 2 appears to stably locomote away in this example, there is no basis in assuming it does so. Similarly, actuation within the frequency bands discussed above would result in the vehicle approaching a stable oscillatory trajectory rather than oscillating about the initial position. In this result, we emphasize the importance of exerting control over a particular agent in a multi-agent setting given that we actuate the platform according to \eqref{eqn:controlrotated}. The ability to actuate the platform in this way leads to questions concerning the control of multiple agents. Naively, one may conclude this control methodology can be targeted to one agent and switched at any time to target another to disperse or station-keep agents as needed, but this is nullified by each vehicle having attained nonzero momentum.
\newpage
\section{Conclusion and Future Work}
\label{section:futurework}
We first developed the reduced equations for a system consisting of a passively compliant Chaplygin beanie atop a translationally compliant platform with finite inertia and proved that for trajectories corresponding to $p_x^2 + p_y^2 = 0$, all of the system's rotational angular momentum is converted into forward translational momentum, resulting in a stable undulatory behavior for arbitrary initial conditions in $\phi$. We then established a characterization for control of a multi-vehicle system coupled dynamically to a medium that gives rise to rich dynamical behavior. Applications of the resulting control methodology to passively compliant multi-agent systems include robotic sorting mechanisms like that of \cite{reznik2001} and the exogenous control of cardiac cells in \cite{Nitsan2016}. 

\begin{figure}
    \centering
    \includegraphics[width=0.4\textwidth,trim={0 0cm 0 0cm},clip]{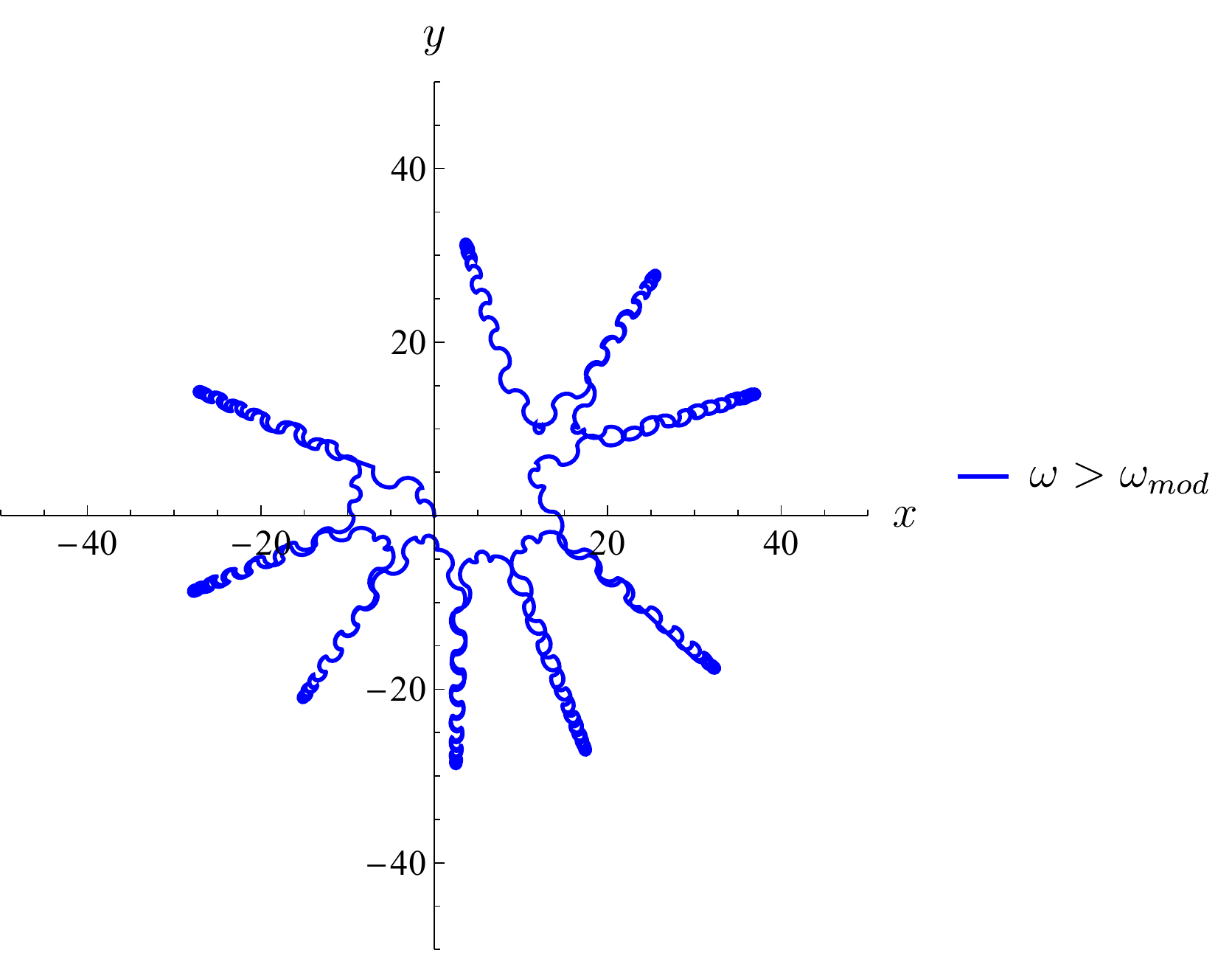}
    \caption{Trajectory resulting from actuating the platform at a frequency of $\omega < \omega_{mod}$ for a Chaplygin beanie with parameters $C = 0.5$, $m = B = k = 1$ for a duration of 500 simulation seconds.}
    \label{fig:trajectoryabovemodal}
\end{figure}

Though it is clear based on the present work that there exist parameter-invariant bounds on platform actuation frequencies for achieving stable undulatory behavior of a passive vehicle under exogenous control, formal proof for this result is sought. Other interesting phenomena are exhibited by the nonlinear dynamics that warrant further exploration. Namely, certain platform actuation frequencies yield behaviors which indicate the presence of multi-scale time dynamics, demonstrated in Fig. \ref{fig:trajectoryabovemodal}. This behavior relates qualitatively to that exhibited by a three-link snake-like robot in \cite{Kelly2016}.
The system was given an initial position at the origin and controlled corresponding to \eqref{eqn:controlrotated}. The vehicle locomotes away along some heading for some time, reverses direction, locomotes for some time, switches its heading, and repeats this behavior. The dynamics of moving along in some heading occur at a fast time scale, while the dynamics for switching direction occur at a much slower time scale. Changes in the direction taken by the vehicle likely correspond to bifurcations in the dynamics at one of these time scales, the analysis of which is a topic of future work. The actuated system has also been shown to display stable oscillatory behavior for frequencies within the frequency bands discussed. As such, proving stabilizability for the controlled system will also be included in future publications.

\bibliographystyle{asmems4}
\begin{acknowledgment}
This work was supported by NSF grant CMMI-1727889. We also thank the reviewers of this work for their comments and suggestions.
\end{acknowledgment}

%
% ...

\bibliography{asme2e}



\end{document}